# Oscillation Frequencies for Simultaneous Trapping of Heteronuclear Alkali Atoms

Kiranpreet Kaur[a], B. K. Sahoo[b] and Bindiya Arora[a*]

[a]*Department of Physics, Guru Nanak Dev University, Amritsar, Punjab-143005, India*
[b]*Theoretical Physics Division, Physical Research Laboratory, Navrangpura, Ahemadabad-380009, India*
e-mail[*] : bindiya.phy@gndu.ac.in

We investigate oscillation frequencies for simultaneous trapping of more than one type of alkali atoms in a common optical lattice. For this purpose, we present numerical results for "magic" trapping conditions, where the oscillation frequencies for two different kind of alkali atoms using laser lights in the wavelength range 500-1200 nm are same. These wavelengths will be of immense interest for studying static and dynamic properties of boson-boson, boson-fermion, fermion-fermion, and boson-boson-boson mixtures involving different isotopes of Li, Na, K, Rb, Cs and Fr alkali atoms. In addition to this, we were also able to locate a magic wavelength around 808.1 nm where all the three Li, K, and Rb atoms are found to be suitable for oscillating at the same frequency in a common optical trap.

## 1 Introduction

Cold atomic and molecular trappings have been the subject of extensive interest for investigating exotic quantum phase transitions, to carry out various high precision measurements, for quantum information experiments etc. [1-4]. This has led to investigate simultaneous production of quantum degenerate Bose-Einstein and Fermi-Dirac gases [5-8]. Realization of bosonic-fermionic atomic mixtures have inspired intensive theoretical and experimental activities to verify various phase separations, influence of the superfluidity of the Bose-Einstein condensate (BEC) on the Fermi degeneracy, dilution refrigeration [9-11] etc. Mixtures of bosonic and fermionic quantum systems, with an eminent example of $^4$He- $^3$He fluids, have attracted intense theoretical and experimental probes on their dynamical behaviour [9] that have provided signatures of many microscopic effects. The mixtures of magnetically-trapped alkali metal atoms such as Na-Cs and Na-K can be used [12] to search for evidence of an electric dipole moment in order to test for parity and time reversal symmetry violations. Besides this, Schloder *et al.* have also investigated the collisional properties of Li and Cs, which are simultaneously confined in a combined magneto-optical trap (MOT), by comprehensively studying trap-loss collisions between the two species [13]. Although optical lattices allow long measurement times with reduced Doppler shifts, however due to ac Stark effect, the trapping fields cause shifts in the internal energy levels of the cold atoms depending on the intensity of the applied lasers. This effect can be eventually nullified by trapping atoms at the magic wavelengths [14]. Magic wavelengths for trapping single species atoms have been reported for a number of atoms including the alkali atoms [14-17]. Simultaneous trapping of two different atomic species have also been reported in Refs. [12,13,18,19], but magic wavelengths of these combined species have not been studied extensively.

In this work, we aim to determine trapping magic wavelengths for the ground states of the Li, Na, K, Rb, Cs, and Fr alkali atoms confined in the same optical trap. The above problem can be resolved by using an optical lattice operating at "magic" trapping conditions (i.e. at the magic wavelengths) for which the oscillation frequency of different trapped atoms remains the same. We have taken into account the wavelength regime, where the primary resonances of the above alkali atoms appear. Since applied external laser fields are arbitrary, the important quantities that are theoretically relevance to find out these conditions are the dynamic dipole polarizabilities. At the magic wavelengths, the ratios of the frequency dependent electric dipole polarizabilities to the atomic weights between the ground states of both the atoms have to be nullified. At the primary resonances, if there are changes in the signs of the dynamic polarizabilities of the atomic states then they can give rise to magic wavelengths. This was demonstrated in Ref. [14], where the frequency-dependent polarizability values of many alkali-metal atoms (Li, Na, K, Rb and Cs) were calculated using an all-order single-double (SD) method, over a range of wavelengths (from the ultraviolet through infrared spectrum). In the present work, we consider these alkali atoms along with the magic trapping conditions for the Fr atoms to look for the magic wavelengths that are common to all these atoms. We employ a relativistic coupled-cluster (RCC) method considering all the non-linear effects at the singles and doubles excitation approximation that is explained in detail in our earlier works [20-22]. The laser cooled and trapped Fr atoms would be useful for studying fundamental symmetry violations such as searching for an electron permanent electric dipole moment and probing atomic parity non-conservation effects [23,24]. It is also one of the most experimentally investigated radioactive elements in terms of atomic laser spectroscopy: many measurements of energy levels [25,26], lifetimes [27-29], isotope shifts [30] and hyperfine structures [20,30] have been reported. Laser cooling and trapping of different Fr isotopes have also been accomplished by different groups [30, 31]. Furthermore, there have been many theoretical calculations of energy levels, dipole matrix elements, lifetimes of states, isotope shifts and hyperfine structure and the enhancement factors for symmetry violation studies [20-22,32-38]. In addition to this, in a certain work presented in [39], the first quantum-degenerate mixture of two different fermionic atomic species and the first triple-degenerate fermion-fermion-boson mixtures were produced. The quantum-degenerate mixtures were also realized using the sympathetic cooling of the fermionic species of $^6$Li and $^{40}$K. Evaporatively cooled gas of bosonic $^{87}$Rb atoms in a MOT are also advantageous in a broad range of possible future experiments, including the creation of heteronuclear fermion-fermion dimers and investigating the BEC and Bardeen-Cooper-Schriffer (BCS) cross-over regime. So keeping this in mind, we intend in this work to determine magic wavelengths for the considered alkali atoms at which more than one species of atoms can be trapped together for carrying out many sophisticated experiments by effectively getting rid of the dominant Stark shifts.

## 2 Theoretical Approach

We We describe briefly here the procedure adopted in the present work for the calculation of the dynamic dipole polarizabilities in the considered alkali atoms. The dynamic polarizability of the ground state $|\Psi_n\rangle$ in an atom is given by

$$\alpha(\omega) = \sum_I \left[ \frac{|\langle\Psi_n|D|\Psi_I\rangle|^2}{E_I-E_n+\omega} + \frac{|\langle\Psi_n|D|\Psi_I\rangle|^2}{E_I-E_n-\omega} \right] = \frac{2}{3(2J_n+1)} \sum_I \frac{(E_I-E_n)|\langle\Psi_n\|D\|\Psi_I\rangle|^2}{(E_I-E_n)^2-\omega^2}, \quad (1)$$

where $J_n = 1/2$ is the total angular momentum of the corresponding ground state, $n$ represents for the principal quantum number, sum over $I$ represents all possible allowed intermediate states due to the dipole operator $D$, $E$'s are the energies of the corresponding states and $\langle\Psi_n\|D\|\Psi_I\rangle$ are the electric dipole (E1) reduced matrix elements between the states $|\Psi_n\rangle$ and $|\Psi_I\rangle$. For the convenience to carry out these frequency dependent polarizabilities of these ground states having a closed core and a valence electron for a wide range of frequencies, we can divide the above expression into various correlation contributions as [40]

$$\alpha = \alpha_c + \alpha_{vc} + \alpha_v, \quad (2)$$

where the notations $c$, $vc$ and $v$ correspond to the core, core-valence and valence contributions respectively. To determine the valence contributions, we employ a sum-over-states approach by evaluating the E1 matrix elements by our RCC method [20-22] and using the experimental energies.

In this RCC approach, we can express $|\Psi_n\rangle$ as

$$|\Psi_n\rangle = e^T\{1+S_n\}|\Phi_n\rangle, \quad (3)$$

where $|\Phi_n\rangle$ is the reference state defined as $|\Phi_n\rangle = a_n^\dagger|\Phi_0\rangle$, with $|\Phi_0\rangle$ being the Dirac-Fock (DF) function for the closed core. Here $T$ and $S_n$ operators account excitations of the electrons from the core orbitals alone and valence orbital together with core orbitals, respectively. We have described this approach to determine amplitudes of the $T$ and $S_n$ excitation operators in our earlier papers [20-22]. Use of the E1 matrix elements along with the experimental energies can give more accurate contributions to the valence correlation contributions of the polarizabilities $\alpha(\omega)$. In fact, many precise E1 matrix elements of the primary transitions in the Na, K and Rb atoms have been compiled in Ref. [41]. So for more accurate determination of the dynamic polarizabilities, we replace our calculated E1 matrix elements when the compiled data from Ref. [41] are found to be more precise. We also use the E1 matrix elements for the $6S-6P$ transitions in the Cs atom reported by Rafac *et al.* [42]. Similarly, few of the E1 matrix elements of the primary transitions in Fr are taken from Ref. [27].

The high-lying excited state (tail) contributions to the valence correlations are comparatively smaller and can be used at the mean-field level approximation using the DF method. Also, the other core and core-valence correlations contributions are insignificant and can be evaluated with sufficient accuracies using simpler many-body methods. Here, we employ random phase approximation (RPA) and a third order relativistic many-body perturbation theory (MBPT method) to determine the core and core-valence contributions, respectively, as discussed in Ref. [43]. Subsequently by calculating the frequency dependent dipole polarizabilities of the Li, Na, K, Rb, Cs and Fr alkali atoms, we plot them against the frequencies to find out the wavelengths $\lambda$ at which at least two species of alkali-metal atoms can have same oscillation frequencies so that they can be trapped in a common optical lattice. These wavelengths are determined by verifying the relation [14]

$$s \equiv \sqrt{\frac{\alpha_1(\lambda)}{m_1}\frac{m_2}{\alpha_2(\lambda)}} \approx 1, \tag{4}$$

where $m_i$ and $\alpha_i(\lambda)$ are the atomic weight and dipole polarizability of the respective alkali atom. So our aim would be to look for the wavelengths $\lambda$'s at which the relation $\alpha_1(\lambda)/m_1 = \alpha_2(\lambda)/m_2$ are almost satisfied.

## 3 Results

Table 1 **Calculated values of the static dipole polarizabilities (in a.u.) for the Li, Na, K, Rb, Cs, and Fr alkali-metal atoms. Our values are compared with the other available theoretical and experimental results. References are given in the square brackets.**

| Contribution | Li | Na | K | Rb | Cs | Fr |
|---|---|---|---|---|---|---|
| $\alpha_v$ | 162.6 | 161.9 | 284.3 | 309.3 | 383.3 | 296.1 |
| $\alpha_c$ | 0.22 | 0.9 | 5.5 | 9.1 | 15.8 | 20.4 |
| $\alpha_{vc}$ | 0 | 0 | -0.13 | -0.26 | -0.47 | -0.95 |
| $\alpha_{tail}$ | 1.2 | 0.08 | 0.06 | 0.11 | 0.19 | 1.26 |
| Total ($\alpha_n$) | 164.1 | 162.4 | 289.8 | 318.3 | 398.8 | 316.8 |
| Others | 164.112 [44] | 162.9 [45] | 289.3 [46] | 315.7 [47] | 399.0 [48] | 317.8 [49] |
| Experiment | 164.2 [50] | 162.1 [51] | 290.58 [52] | 318.79 [52] | 401.0 [53] | |

In Table 1, we present the static dipole polarizabilities of the alkali-metal atoms, along with the detailed breakdown of the various contributions and compare them with the earlier theoretical and experimental results. The details of these calculations are also presented in Ref. [15]. Comparison of these values with their corresponding experimental results will ascertain about the accuracy in the results. This can also ensure about similar accuracy in the estimated dynamic polarizabilities. The most accurate experimental measurement of Li ground-state polarizability has been reported as $\alpha_{2s} = 164.2$ in atomic unit (a.u.) [50]. Our result $\alpha_{2s} = 164.1$ a.u. is in excellent agreement with this experimental value. The most stringent experimental value for Na ground-state polarizability was obtained by interferometry experiments as $\alpha_{3s} = 162.1$ a.u. [51], and our present value $\alpha_{3s} = 162.4$ a.u. agrees well with the experimental value. The experimental result available for the ground-state polarizability of K is $\alpha_{4s} = 290.58$ a.u. [52], which is very close to our calculated value $\alpha_{4s} = 289.8$ a.u. In the same table, we also list the polarizability of the 5$S$ state of Rb as 318.3 a.u, while the most precise experimental result reported is 318.79 a.u. [52]. This is also in excellent agreement with our result. Our calculated value of the static polarizability for the 6$S$ state of Cs atom is 398.8 a.u., which agrees well with the value 399.0 a.u. obtained by Borschevsky *et al.* [48] using another RCC method and experimentally measured value 401.0 a.u. of Amini *et al.* [53] using the time of flight technique. Similarly, our calculated value of static polarizability of Fr atom is 316.8 a.u. is close to the previously reported value as 317.8 a.u. [49] using high accuracy experimental data of the E1 matrix elements. Thus from the comparison between the other available values and out calculated results, as shown in the table, it is clear that our static polarizabilities are very accurate.

To find out the magic wavelengths $\lambda$, we plot the dynamic polarizabilities of the ground states of all the alkali atoms and they are found at the crossings of curves for the ratio of polarizability to the atomic weight for the corresponding alkali atoms. We particularly choose atomic weights $m_i$ for the $^7$Li, $^{23}$Na, $^{40}$K, $^{85}$Rb, $^{133}$Cs, and $^{223}$Fr alkali atoms as 7.016003 a.u., 22.989769 a.u., 39.96399 a.u., 84.911794 a.u, 132.90542 a.u. and 223.0197 a.u. respectively from NIST database. The reason for considering these particular isotopes is that they are often used in the cold atom experiments. In Fig. 1, we show the ratios of the above isotopes as function of wavelength λ. In this figure, the red, green, blue, pink, neon blue and yellow coloured curves corresponds to the ⁷Li, ²³Na, ⁴⁰K, ⁸⁵Rb, ¹³³Cs, and ²²³Fr atoms respectively. As a practice, the magic wavelengths which are extremely close to the resonances are not taken into account. We also found that the width of the matches in the wavelength range of 300-500 nm is very narrow and their corresponding polarizability values are also small. Hence, we have omitted all possible matches at wavelengths below 500 nm. Similarly, we have omitted the matches at the higher wavelengths above 1200 nm that are adjoining to the resonances. From the figure, we can see that there is one magic wavelength at 549.09 nm for the $^7$Li and $^{23}$Na atoms and this supports for a blue or a dark detuned trap for simultaneous optical trapping. Similarly, for a pair of $^7$Li and $^{40}$K atoms, magic wavelength λ is located at 808.02 nm and for the $^7$Li and $^{85}$Rb atoms, we find there are two magic wavelengths at 786.79 nm and 807.92 nm supporting red detuned traps. We also find magic wavelengths for the $^7$Li and $^{133}$Cs atoms around 864 nm and 907 nm, both supporting the red detuned traps. Magic wavelengths at 720.12 nm and 820.49 nm occur for the simultaneous trapping of the $^7$Li and $^{223}$Fr atoms with effectively null differential Stark shifts. Similarly, at least two magic wavelengths were located for the other combinations of the alkali-metal atoms. In the same figure, we can also clearly observe that simultaneous trapping of the $^7$Li, $^{40}$K and $^{85}$Rb atoms are possible at the magic wavelength of 808.13 nm. Using these species, it would be interesting to study dynamics of a boson-fermion-boson mixture.

Table 2 **Wavelengths $\lambda$ in nm, at which a pair of alkali atoms among ⁷Li, ²³Na, ⁴⁰K, ⁸⁵Rb, ¹³³Cs, and ²²³Fr can have the same frequencies of oscillation at which a common optical trap can be used for trapping them. The corresponding ground state frequency dependent polarizabilities $\alpha_1$ and $\alpha_2$ (in a.u.) for an 1-2 atom pair are also given.**

| Alkali Pair (1-2) | Wavelength $\lambda$ (nm) | $\alpha_1$ | $\alpha_2$ |
|---|---|---|---|
| ⁷Li – ²³Na | 549.09 | -329.38 | -1078.2 |
| ⁷Li – ⁴⁰K | 808.02 | 522.15 | 2972.2 |
| ⁷Li – ⁸⁵Rb | 786.79 | 594.66 | 7192.7 |
|  | 807.92 | 523.41 | 6330.8 |
| ⁷Li – ¹³³Cs | 864.24 | 412.92 | 7817.18 |
|  | 907.99 | 358.86 | 6793.8 |
| ⁷Li – ²²³Fr | 720.12 | 1238.8 | 39356.3 |
|  | 820.49 | 491.61 | 15618.1 |
| ²³Na – ⁴⁰K | 769.02 | 321.95 | 559.84 |
| ²³Na – ⁸⁵Rb | 789.61 | 370.21 | 1367.9 |
|  | 950.07 | 263.81 | 974.77 |
| ²³Na – ¹³³Cs | 875.95 | 296.44 | 1714.41 |
|  | 1020.07 | 245.89 | 1422.03 |
| ²³Na – ²²³Fr | 732.09 | 461.67 | 4480.47 |
|  | 836.32 | 324.25 | 3146.8 |
| ⁴⁰K – ⁸⁵Rb | 784.59 | 6901.9 | 14665.7 |
|  | 808.02 | 2965.4 | 6301.2 |
| ⁴⁰K – ¹³³Cs | 769.02 | -447.55 | -1488.5 |
|  | 869.39 | 1315.1 | 4373.7 |
|  | 938.03 | 874.32 | 2907.9 |
| ⁴⁰K – ²²³Fr | 710.69 | 1679.1 | 9371.3 |
|  | 768.16 | 83.92 | 468.34 |
|  | 821.35 | 2238.6 | 12493.6 |
| ⁸⁵Rb – ¹³³Cs | 789.61 | -1249.0 | -1954.9 |
|  | 873.68 | 1605.7 | 2513.1 |
|  | 1130.2 | 606.26 | 948.91 |
| ⁸⁵Rb – ²²³Fr | 694.61 | -1102.1 | -2894.8 |
|  | 790.47 | -178.31 | -468.34 |
|  | 822.21 | 3627.3 | 9527.4 |
| ¹³³Cs – ²²³Fr | 616.13 | -373.45 | -626.69 |
|  | 809.34 | -2885.3 | -4841.8 |
|  | 878.82 | 744.24 | 1248.91 |

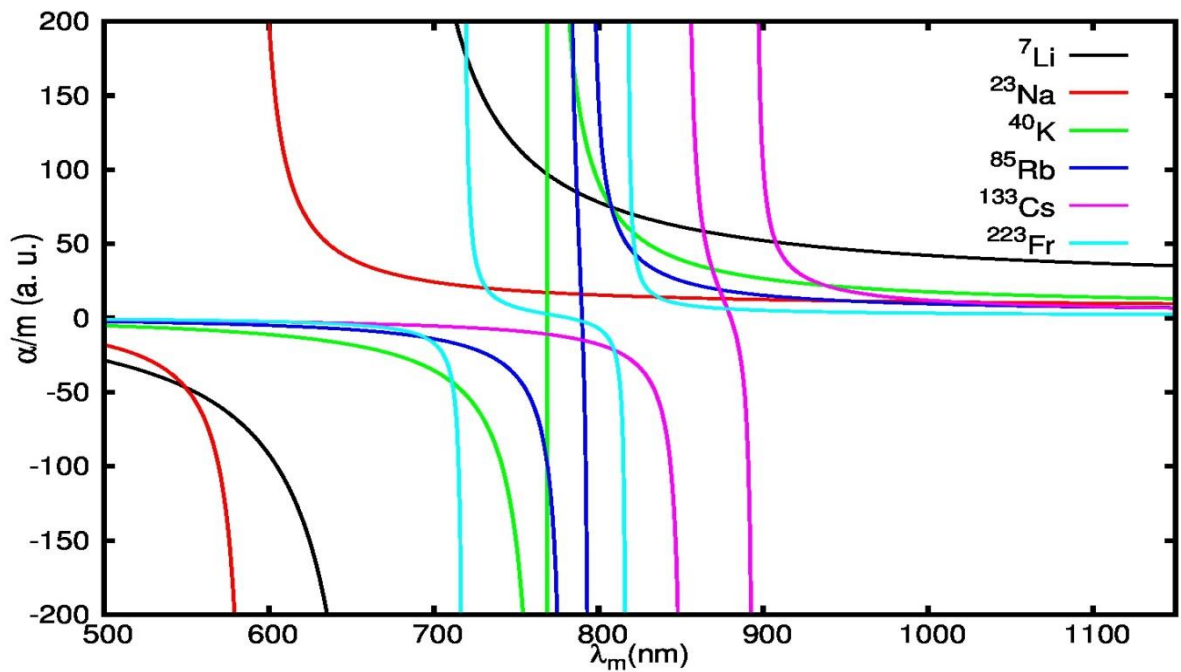

**Figure 1.** The ratios of frequency dependent polarizabilities to the atomic weights $\alpha(\lambda)/m$ of the $^7$Li, $^{23}$Na, $^{40}$K, $^{85}$Rb, $^{133}$Cs, and $^{223}$Fr alkali atoms as functions of wavelength λ (in nm).

The wavelengths for which Eq. (4) is satisfied are tabulated in Table 2 for different combinations of a pair of species for optically trapping of $^7$Li, $^{23}$Na, $^{40}$K, $^{85}$Rb, $^{133}$Cs, and $^{223}$Fr alkali atoms oscillating with the same frequency. It is worth mentioning here that all wavelengths are given in vacuum. It can be clearly seen in the table that we get a set of 32 magic wavelengths for all the combinations of the considered alkali atoms in the wavelength range 600-1200 nm. Out of these thirty two, seven wavelengths supports blue detuned traps while the rest of the wavelengths are supporting the red detuned traps. These seven magic wavelengths supporting blue or dark detuned traps includes 549.09 nm for the pair $^7$Li – $^{23}$Na, 769.02 nm for the pair $^{40}$K – $^{133}$Cs, 789.61 for the pair $^{85}$Rb – $^{133}$Cs, 694.61 nm and 790.47 nm for the pair $^{85}$Rb – $^{223}$Fr, 616.13 nm and 809.34 nm for the pair $^{133}$Cs – $^{223}$Fr

**4 Conclusion**

In the foregoing work, we have investigated the ratios of the frequency dependent electric dipole polarizabilities to the atomic weights of the alkali atoms in the infrared spectral region. The electric dipole polarizabilities were evaluated by calculating the important electric dipole matrix elements using a relativistic coupled-cluster method. The non-dominant core and core-valence correlation contributions were estimated by the random phase approximation and a third order relativistic many-body perturbation theory respectively. The net values of the polarizabilities are compared with the available experimental and theoretical results and are found to be in good agreement. By plotting the ratios of the frequency dependent electric dipole polarizabilities to the atomic weights against wavelength, the oscillation frequencies for the alkali atoms are identified at which their ground states can experience same amount of Stark shifts when they are trapped simultaneously in a common optical lattice. This knowledge of trapping simultaneously heteronuclear alkali atoms with the effectively null differential stark shifts would be of immense interest to the experimentalists for carrying out many high precision experiments using these atoms; especially in the context of studying dynamic properties of bosonic and fermionic species mixtures.

**Acknowledgement**

The work of B.A. is supported by Council of Science and Industrial Research (CSIR) grant no. 03(1268)/13/EMR-II, India. K.K. acknowledges the financial support from Department of Science and Technology (DST) letter no. DST/INSPIRE Fellowship/2013/758). B.K.S acknowledges use of Vikram-100 HPC Cluster at Physical Research Laboratory, Ahmedabad.